# An entertaining physics: On the possibility of energy storage enhancement in electric capacitors using the compensational inductive electric field.

Alexander Khitun

Department of Electrical and Computer Engineering, University of California – Riverside, Riverside, California, 92521, USA.

**Abstract**

In this work, we consider the possibility of energy storage enhancement in electric capacitors using the compensational method. The essence of the proposed approach is the use of inductive voltage $V_{ind}$ to partially compensate the electrostatic voltage $q/C$ produced by the electric charges on the capacitor plates. We hypothesize that it may be possible to increase the amount of charge stored on the plates before the breakdown and increase the energy stored in the capacitor using the compensational inductive voltage. There are several possible scenarios of manipulating the inductive voltage to increase the amount of energy released via the discharge. We also consider several electro-magnetic capacitors for practical utilization. Potentially, the energy per volume stored in a simple parallel plate capacitor may exceed the one of gasoline. The physical limits and technological shortcomings of the proposed approach are also discussed.

Corresponding Author: akhitun@engr.ucr.edu

## I. Introduction

There is a big impetus in the development of electric energy storage devices which is stimulated by the urgent need in clean (pollution-free) energy sources [1-6]. It would of great benefit to the society to have an electric energy storage device with the energy density (i.e., energy per volume [J/m$^3$]) exceeding the one of gasoline 34.2 MJ/L.

Electric capacitors are among the promising devices offering a convenient way for energy storage and release [7]. Recently, the most of research has been focused on supercapacitors [8-11]. Here, we want to reconsider some basic derivations on the energy stored in classical capacitors (e.g., a parallel plate capacitor) and offer a novel approach to energy storage enhancement.

The very well-known formula, which can be found in all textbooks, states that the energy stored in an electric capacitor is given by

$$W = \frac{1}{2}CV^2, \tag{1}$$

where $C$ is the capacitance and $V$ is the voltage difference across the plate. The maximum amount of energy stored in any given capacitor $W_{max}$ is limited by the breakdown voltage $V_b$ as follows:

$$W_{max} = \int_0^{q_{max}} V dq = \int_0^{q_{max}} \frac{q}{C} dq = \frac{q_{max}^2}{2C} = \frac{1}{2}CV_b^2, \tag{2}$$

where $q_{max}$ is the maximum amount of charge stored on the plates. *We want to emphasize that these formulas (1-2) are derived under the assumption that the voltage difference between the plate is solely due to the electric charge $V = q/C$. In this work, we consider the possibility of using* compensational inductive voltage $V_{ind}$ *for energy storage enhancement.*

### II.     Principle of operation

According to Eq.(2), the maximum energy stored in the given capacitor is limited by the breakdown voltage $V_b$. The essence of the proposed compensational approach is the use the inductive voltage $V_{ind}$ to partially compensate the electrostatic voltage $q/C$ produced by the electric charges and accumulate a larger amount of charge on the plates. Then, it may be possible to increase the amount of energy released during the discharge by synchronizing the changes in the electrostatic and inductive voltages.

To explain this idea, let us consider a parallel plate capacitor placed in the time-varying magnetic field as shown in Fig.1. It is a parallel plate capacitor with two conducting plates of area A oriented in X-Z plate. The separation distance between the plates is *d*. For simplicity, we consider the volume between the plates filled with vacuum. Each plate carries a charge of magnitude *q*. There is a source of time-varying magnetic field *B* directed along/opposite the Z axis. The electric field between the plates $\vec{E}_{eff}$ is a superposition of two:

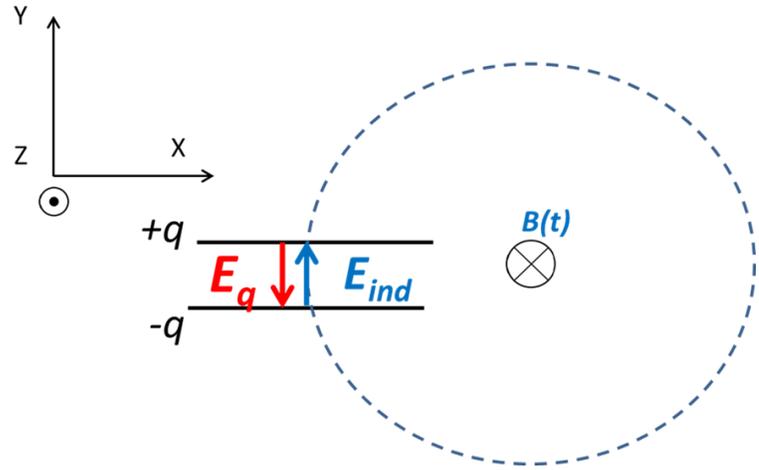

**Figure 1:** Schematics of a parallel-plate capacitor in a time-varying magnetic field *B*. The red and the blue arrows show the direction of the electrostatic $\vec{E}_q$ and the inductive electric field $\vec{E}_{ind}$, respectively. The direction of the inductive field is chosen to compensate the electrostatic field.

$$\vec{E}_{eff} = \vec{E}_q + \vec{E}_{ind}, \qquad (3)$$

where $\vec{E}_q$ is the electrostatic field produced by the electric charges on the plates, $\vec{E}_{ind}$ is the inductive electric field produced by the time-varying magnetic field. These two electric fields are completely independent on each other and defined by the different physical quantities. According to Gauss's law for electric field, the strength of the electric field $\vec{E}_q$ produced by electric charges can be found as follows:

$$\nabla \cdot \vec{E}_q = \frac{\rho}{\varepsilon_0}, \qquad (4)$$

where $\rho$ is the charge density, $\varepsilon_0$ is the permittivity of free space. The strength of the inductive electric field produced by the time-varying magnetic field is given by the Faraday's law of induction:

$$\nabla \times \vec{E}_{ind} = \frac{-\partial \vec{B}}{dt} \qquad (5)$$

where $B$ is the magnetic flux density, $t$ is the time. In our consideration, we take the direction of the inductive electric field to be opposite to the electrostatic field as depicted in Fig.1. Hereafter, we call this field *the compensational field*.

*Hypothesis 1:* The application of the compensational inductive field increases the maximum amount of charge which can be stored on the plates before the breakdown.

A proof for Hypothesis 1 is following from the voltage definition. The voltage difference between some points $a$ and $b$ is given by

$$\Delta V_{ab} = -\int_a^b \vec{E} \cdot d\vec{l} = -\int_a^b [\vec{E}_q + \vec{E}_{ind}] \cdot d\vec{l} \qquad (6)$$

The maximum amount of charge which can be stored before the breakdown $q_{max}$ can be estimated from Eq. (6) as follows:

$$\left|\frac{q_{max}}{C} - V_{ind}\right| = V_b, \qquad \rightarrow \qquad q_{max} = (V_b + V_{ind}) \cdot C \qquad (7)$$

According to Eq. (7), there is no physical limit for $q_{max}$ as long as the total voltage difference between the plates of the capacitor is less than the breakdown voltage.

The total energy stored in the capacitor shown in Fig.1 includes electric and magnetic parts. A complete energy analysis should include the capacitor and the source of magnetic field as well. Our primary interest is focused on the amount of electric energy that can be released by discharging the capacitor. The increase of the maximum amount of charge $q_{max}$ itself does not guarantee any increases in the energy released. For instance, the energy released at the discharge approaches zero when the electrostatic voltage is completely compensated by the inductive voltage. In this part, we consider possible procedures for inductive voltage synchronization with electrostatic voltage leading to the most effective energy release.

*Hypothesis 2:* The amount of electric energy released from the capacitor using the compensational method may be large than the energy released under the ordinary conditions.

Let us consider an electric capacitor moving in constant magnetic field *B* as shown in Fig.2. We intentionally switch from the capacitor shown in Fig.1 which includes an external source of electric field to the capacitor moving in the constant magnetic field to avoid the analysis of energy transfer at the external source. Our primary interest is to see how much electric energy can be released *locally* in the capacitor. There is the Lorentz force acting on the charge on the plates:

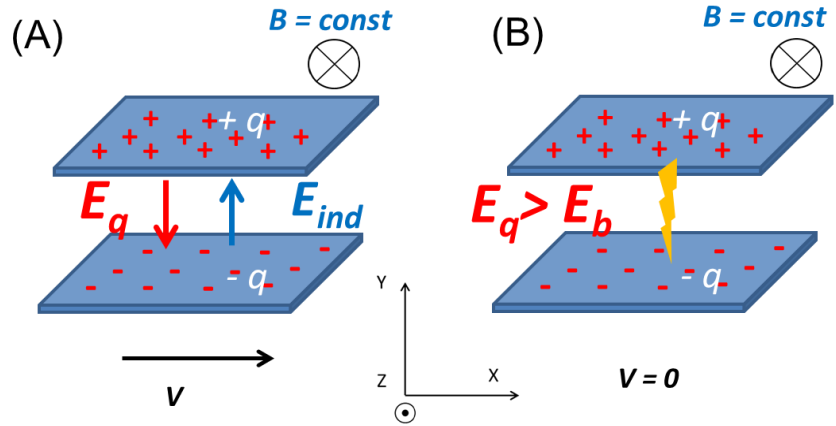

**Figure 2:** Illustration of using the magnetic part of the Lorentz force for compensating the force of electrostatic attraction. (A) A parallel plate capacitor is moving with constant velocity $v > 0$ in the constant and uniform magnetic field $B$. The direction of motion is chosen such that the magnetic part of the Lorentz force compensates the force of electrostatic attraction. The maximum amount of charge stored on the plates may exceed the maximum value in regular conditions. (B) The capacitor is instantly stopped ($v = 0$). The electric field across the plates is defined by the charge only.

$$F = q[E + v \times B], \tag{8}$$

where $v$ is the velocity. The direction of motion is chosen in such a way so the magnetic component of the Lorentz force opposes the electric component. The effective electric field acting on the charges $\vec{E}_{eff}$ is given by

$$\vec{E}_{eff} = \frac{F}{q} = E_x - v_y B_z. \tag{9}$$

In this scenario, the magnetic component prevents the electric breakdown and allows to accumulate an additional amount of charge

$$\delta q = v_y B_z \cdot d \cdot C \qquad (10)$$

*Scenario 1:* The moving capacitor is instantly stopped. The instant stopping means that the time of velocity drops to zero is much shorter compared to the time required for electric discharge (e.g. the time for electrons to move the distance between the plates). In this scenario, the energy of the stopped capacitor is given by the classical formula Eq.(1). However, the maximum amount of charge is not limited by the breakdown voltage $V_b$ anymore. The energy in the stopped capacitor is solely determined by the electrostatic electric field as:

$$W_{max} = \tfrac{1}{2} C (V_b + V_{ind})^2, \qquad (11)$$

where $V_{ind} = v_y B_z d$ is the compensating inductive voltage. In scenario 1, the release of the stored energy has a form of an irreversible electric breakdown.

*Scenario 2:* The energy stored can be released in a reversible manner by synchronizing the electric discharge with the change of the compensational electric field. We consider three stages of capacitor operation including charging, charge storage, and discharging as illustrated in Fig.3. The graph shows the change of electric field in tine. The green and the blue curves correspond to the electrostatic and inductive fields, respectively. The red curve depicts the effective electric field between the plates.

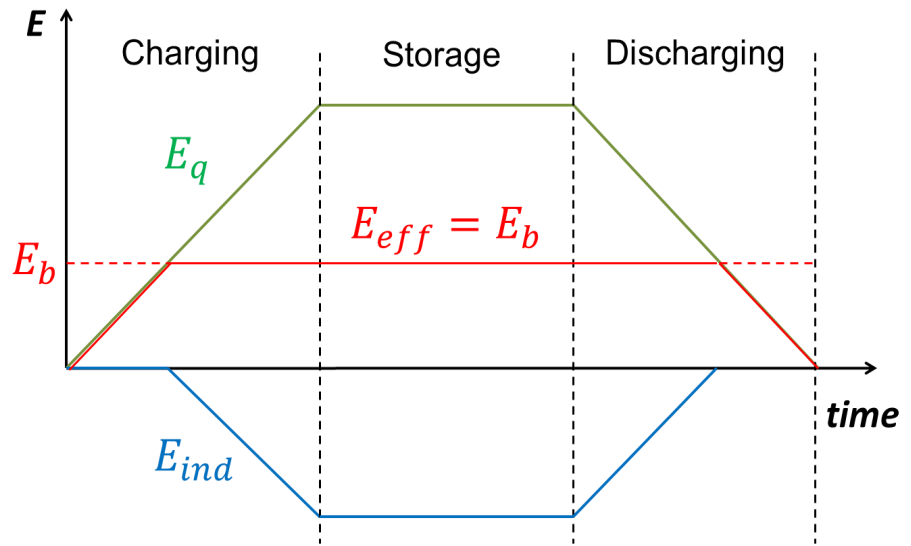

**Figure 3:** Illustration of the principle of operation. The charging of the capacitor starts at zero inductive electric field $E_{ind}$. As the electric field produced by charges $E_q$ approaches the breakdown value $E_b$, the source of magnetic field is turned on to provide the inductive electric field $E_{ind}$. The inductive field $E_{ind}$ is directed opposite to $E_q$ to compensate the charge-induced electric field. The charging is continued till some charge is stored on the plates. In the storage regime, the inductive field $E_{ind}$ remains constant to keep the effective electric field below the breakdown. The discharging of the capacitor is synchronized with the change of the magnetic field rate. The effective magnetic field is kept close to $E_b$ till the last electron is discharged.

*Charging.* It starts at the zero compensational field ($\vec{E}_{ind} = 0$). As the electric field produced by charges $E_q$ approaches the breakdown value $E_b$, the source of magnetic field is turned on to provide the compensational electric field $E_{ind}$ in the direction opposite to $E_{in}$ to meet the below breakdown requirement Eq.(7). The charging is continued till some charge $q_{max} = d \cdot C \cdot (E_b + E_{ind})$ is stored on the plates.

*Storage.* In order to store the accumulated amount of charge, the compensational field should remain constant to keep the effective electric field below the breakdown according to Eq. (7).

*Discharging.* The discharging of the capacitor is synchronized with the change of the inductive electric field to keep the effective electric field just below the $E_b$. The energy released during the discharging is

$$W_{max} = \int_0^{q_{max}} V dq = \frac{1}{2} C V_b^2 + C(V_{ind} V_b). \tag{12}$$

Comparing Eq.(2) and Eq.(12), one can see the fundamental difference. *The maximum energy stored in the capacitor is no longer limited by the breakdown voltage but depends on the practically achievable magnitude of the compensational field.*

### III. Discussion

In theory, there is no limit to the energy stored in a single capacitor using the compensational method. In practice, the maximum amount of energy and the storage time is limited by the capabilities of the magnetic field source. It may be problematic to keep constant the rate of magnetic field change $\partial B/\partial t$ for a sufficiently long time. As a possible solution, we consider an oscillatory circuit where the charge on the plates of the capacitor and the compensational inductive field oscillates in time. In Fig.4, it is schematically depicted a circuit comprising two complementary electro-magnetic devices. Each device is shown to have four parallel plate capacitors placed around a source of magnetic field *B*. The rate of magnetic field change is adjusted to compensate for the electrostatic force of attraction in the capacitors. The oscillation of the charge and the magnetic field are synchronized to keep the effective electric field below the breakdown. The access charge $\delta q$ (i.e., the charge exceeding the breakdown amount $q_b$ in normal conditions) is transferred between the two complementary devices. The charge on the plates and the external magnetic field oscillate out of phase:

$$q = q_b + \delta q \sin(\omega t) \quad \text{and} \quad B = B_{max} \cos(\omega t), \tag{13}$$

where $\omega$ is the frequency of oscillation. The magnitude of the inductive electric field $B_{max}$ is adjusted to satisfy Eq. (7). The oscillation implies that the energy stored in the capacitor will be transferred from one device to another capacitor via a magnetic field. It is interesting to note that there is no "breakdown magnetic field" as there are no magnetic charges. In order to release the energy stored, the two devices are discharged according to the two possible scenarios as described before.

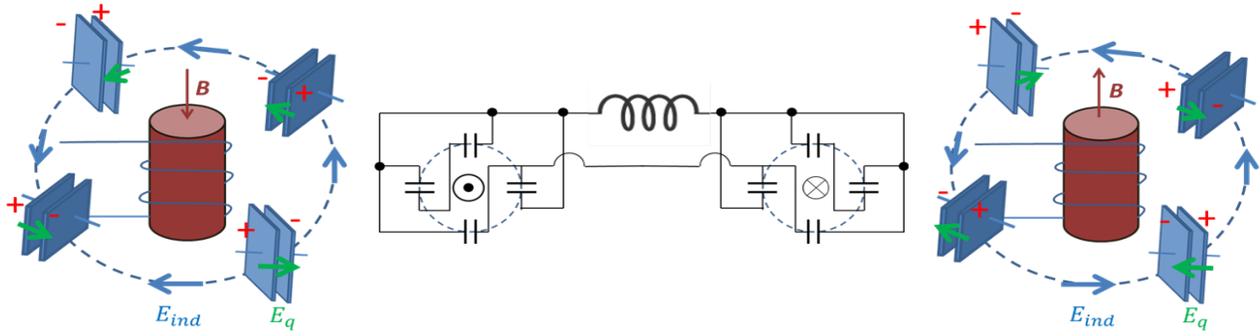

**Figure 4:** Schematics of a complementary electro-magnetic device comprising two coupled oscillators. Each oscillator is shown to have four parallel plate capacitors placed around a source of time varying magnetic field $B$. The rate of magnetic field change is adjusted to compensate the electrostatic force of attraction in the capacitors. The oscillation of the charge and the magnetic field are synchronized to keep the effective electric field below the breakdown.

In order to compete with gasoline, the electrostatic field (i.e. produced by the charges only in the "instantly stopped capacitor") should exceed 1 GV/m, which would require a large-magnitude compensational inductive field. In Fig.5, there are depicted two possible approaches to maximize the inductive voltage. One possibility is to place a capacitor inside the source of time-varying magnetic field. In Fig. 5(A), it is schematically shown a transformer-like structure with parallel-plate capacitors placed inside a non-conducting ferrite core (e.g. $Y_3Fe_2(FeO_4)_3$). The time-varying magnetic field is produced by the alternating current in the primary coil. The inductive voltage may exceed tens of kV/cm at ferromagnetic resonance [12]. Arranged in the complementary circuit (see Fig.4), it may be possible to build an oscillator circuit where the inductive voltage provided by the primary coil compensates the electrostatic voltage preventing the capacitors from breakdown. In Fig.5(B), there are shown the results of numerical modeling depicting the changing in time electrostatic and inductive electric filed. The sum of two does not exceed the breakdown value $E_b$ though the each of the components may be much larger than $E_b$. Power dissipation during storage is the major shortcoming of this approach.

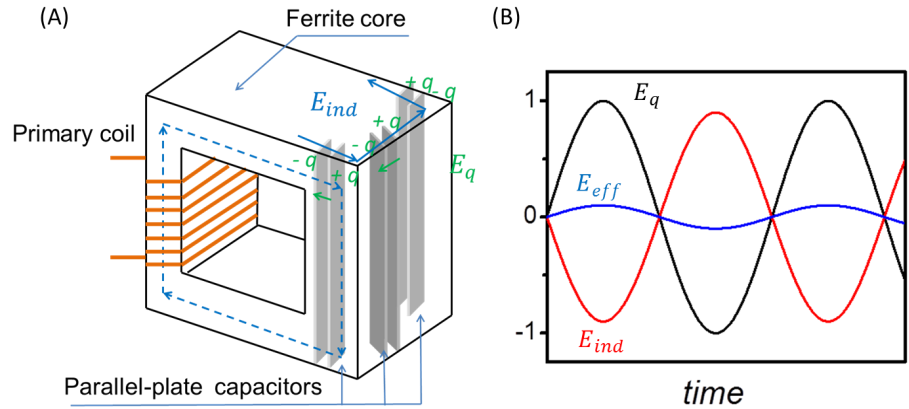

**Figure 5:** (A) Schematics of the proposed electro-magnetic capacitor 1. There are shown three parallel-plate capacitors placed inside a ferrite core. The compensational inductive filed is provided by AC in the primary coil. (B) Results of numerical modeling showing the electrostatic (black curve) and inductive (red curve) fields oscillating in time. The total effective field is below the breakdown value.

To minimize power dissipation (i.e., the leakage electric current between the plates), we consider a device shown in Fig.6(A). A parallel-plate capacitor is placed inside the secondary coil of a transformer. There is a part of the coil made of a semiconductor material (e.g. silicon). The two plates are the sides of the metallic wires connecting the semiconductor. The charge on the sides is provided by the external circuit (not shown in the figure). The secondary coil is an open circuit. There are two metallic spheres (sphere capacitors) on the edges. The equivalent electric circuit is shown in Fig.6(B). It consists of a parallel-plate capacitor in the center, two step up transformers, and two sphere capacitors on the sides. There is infinite resistance between the side sphere capacitors. The transformers are aimed to provide the compensational inductive voltage. Let us consider a steady-state condition, where the inductive voltages produced by the transformers completely compensates the voltage produced by the charges across the plates. The total electric field across the plates of the parallel-plate capacitor is zero:

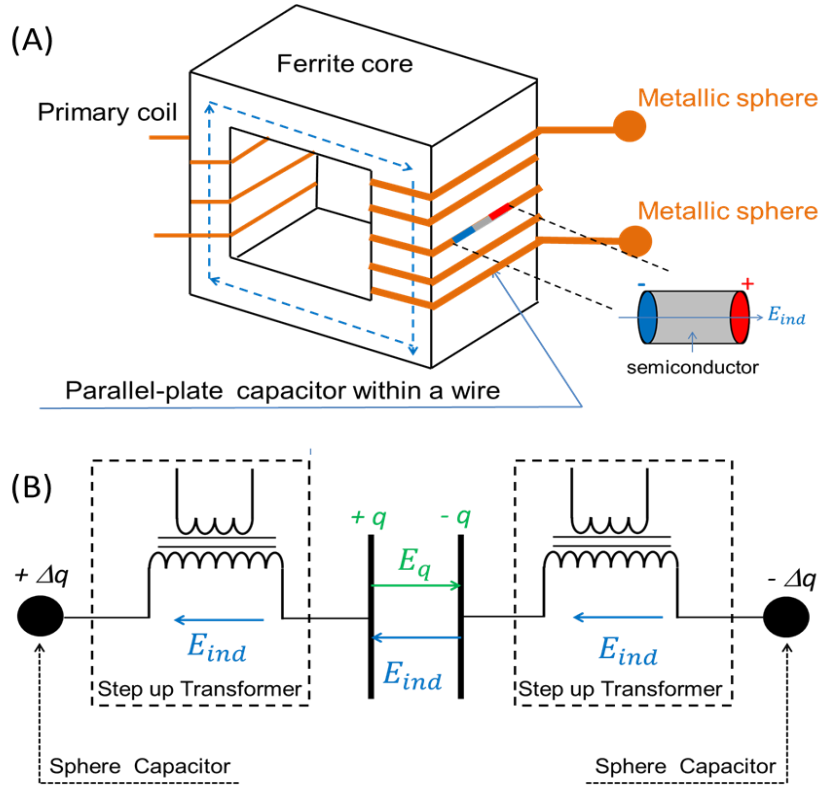

Figure 6: (A) Schematics of the proposed electro-magnetic capacitor 2 where the parallel-plate capacitors are placed inside the secondary coil of the transformer. The capacitor is formed by replacing a part of the coil by a semiconductor. (B) The equivalent circuit. It is an open circuit comprising a parallel-plate capacitor (in the center), two step-up transformers, and two sphere capacitors at the edges. The inductive voltage completely compensates for the electrostatic potential difference between the plates. There is zero leakage current in the secondary coil.

$$V = \frac{q}{C} + 2V_{ind} = 0 \qquad (14)$$

There is no leakage current between the plates. The application of the inductive voltage leads to the redistribution of the electric charge within the secondary coil (e.g., between the parallel-plate capacitor and the sphere capacitors on the sides). The amount of charge $\Delta q$ moved to the sphere capacitors is given by

$$\Delta q / 4\pi\varepsilon_0 r = V_{ind} \qquad (15)$$

where $r$ is the radius of the sphere. The utilization of the step up transformers is one of the possibilities to achieve a large-amplitude compensational field. Though there is no electric current in the secondary coil, the electric current must flow in the primary coil to provide the inductive voltage.

There may be other solutions (e.g., electro-magnetic circuits, charging/discharging procedures, approaches to the inductive electric field generation, etc.) for more efficient energy storage/release. In this work, we want to outline the idea of using the compensational inductive field for energy storage enhancement.

**IV. Conclusions**

It may be possible to increase electric energy stored in electric capacitors by applying a compensational inductive field. This compensational field is not aimed to produce any work but to keep the charges on the plates from breakdown. We hypothesize that the application of the compensational inductive field may increase the maximum amount of charge which can be accumulated on the plates of the electric capacitor before the breakdown. We also considered possible charging /discharging scenarios to maximize energy storage/release by manipulating the compensational field. The major shortcomings of the proposed compensational method are associated with the need for a source of inductive voltage (i.e. a source of time-varying magnetic field) capable to provide a large-magnitude inductive field for a long (e.g., days) period of time. We propose several types of modified electro-magnetic energy storage devices combining electric capacitors with the source of time-varying magnetic field.

There are many questions regarding the inductive field to be clarified. For instance, the effective electric field in the capacitor shown in Fig.2.depends on the velocity of the dielectric as the Lorentz force acts only on the moving charges. It is interesting to analyze the theoretical limit for the capacitor with vacuum between the plates. Is it limited by the vacuum dielectric strength or does not have any limit et al? Another question is related to the inductive voltage distribution within the semiconductor part of the secondary coil (see Fig.6).  What is the practical limit for the compensational inductive field in this case? The key question is related to the feasibility of enhancing electric energy density above the one of gasoline. That would be a breakthrough for many practical applications. The examples described in this work show the entertaining side of applied physics, where old and well-known concepts can be reconsidered and extended for use in the emerging field of energy storage.

**Figure Captions**

**Figure 1:** Schematics of a parallel-plate capacitor in a time-varying magnetic field *B*. The red and the blue arrows show the direction of the electrostatic $\vec{E}_q$ and the inductive electric field $\vec{E}_{ind}$, respectively. The direction of the inductive field is chosen to compensate the electrostatic field.

**Figure 2:** Illustration of using the magnetic part of the Lorentz force for compensating the force of electrostatic attraction. (A) A parallel plate capacitor is moving with constant velocity $v > 0$ in the constant and uniform magnetic field $B$. The direction of motion is chosen such that the magnetic part of the Lorentz force compensates the force of electrostatic attraction. The maximum amount of charge stored on the plates may exceed the maximum value in regular conditions. (B) The capacitor is instantly stopped ($v = 0$). The electric field across the plates is defined by the charge only.

**Figure 3:** Illustration of the principle of operation. The charging of the capacitor starts at zero inductive electric field $E_{ind}$. As the electric field produced by charges $E_q$ approaches the breakdown value $E_b$, the source of magnetic field is turned on to provide the inductive electric field $E_{ind}$. The inductive field $E_{ind}$ is directed opposite to $E_q$ to compensate the charge-induced electric field. The charging is continued till some charge is stored on the plates. In the storage regime, the inductive field $E_{ind}$ remains constant to keep the effective electric field below the breakdown. The discharging of the capacitor is synchronized with the change of the magnetic field rate. The effective magnetic field is kept close to $E_b$ till the last electron is discharged.

**Figure 4:** Schematics of a complementary electro-magnetic device comprising two coupled oscillators. Each oscillator is shown to have four parallel plate capacitors placed around a source of time varying magnetic field $B$. The rate of magnetic field change is adjusted to compensate the electrostatic force of attraction in the capacitors. The oscillation of the charge and the magnetic field are synchronized to keep the effective electric field below the breakdown.

**Figure 5:** (A) Schematics of the proposed electro-magnetic capacitor 1. There are shown three parallel-plate capacitors placed inside a ferrite core. The compensational inductive filed is provided by AC in the primary coil. (B) Results of numerical modeling showing the electrostatic (black curve) and inductive (red curve) fields oscillating in time. The total effective field is below the breakdown value.

**Figure 6:** (A) Schematics of the proposed electro-magnetic capacitor 2 where the parallel-plate capacitors are placed inside the secondary coil of the transformer. The capacitor is formed by replacing a part of the coil by a semiconductor. (B) The equivalent circuit. It is an open circuit comprising a parallel-plate capacitor (in the center), two step-up transformers, and two sphere capacitors at the edges. The inductive voltage completely compensates for the electrostatic potential difference between the plates. There is zero leakage current in the secondary coil.

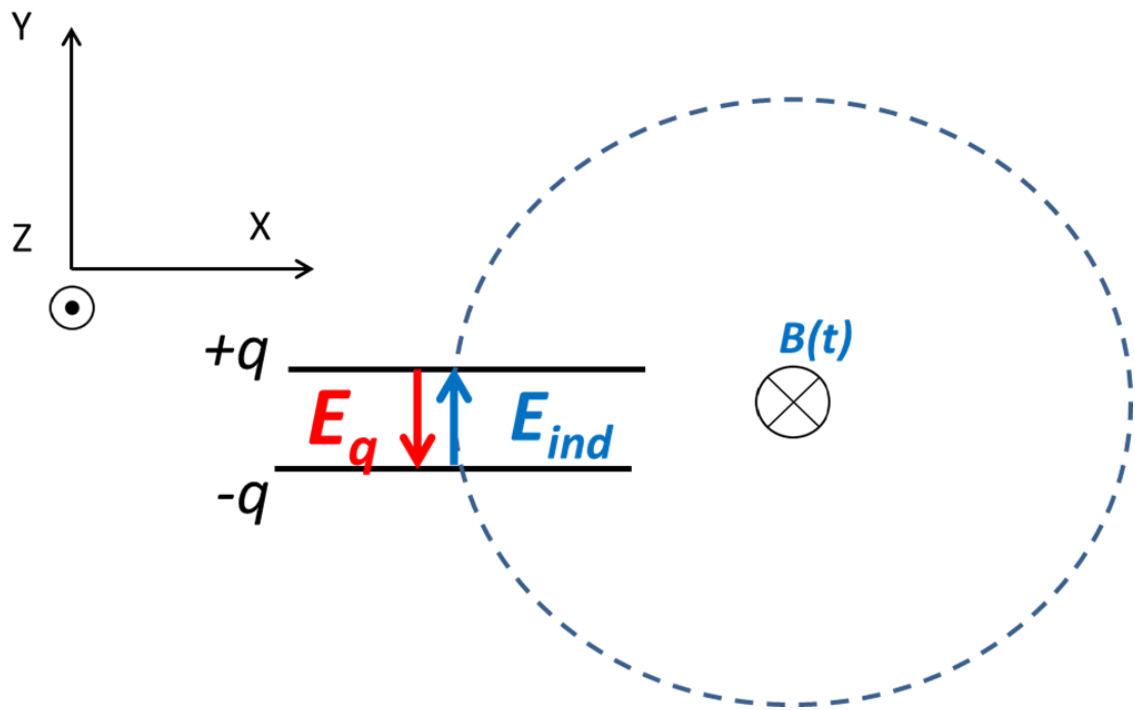

**Figure 1**

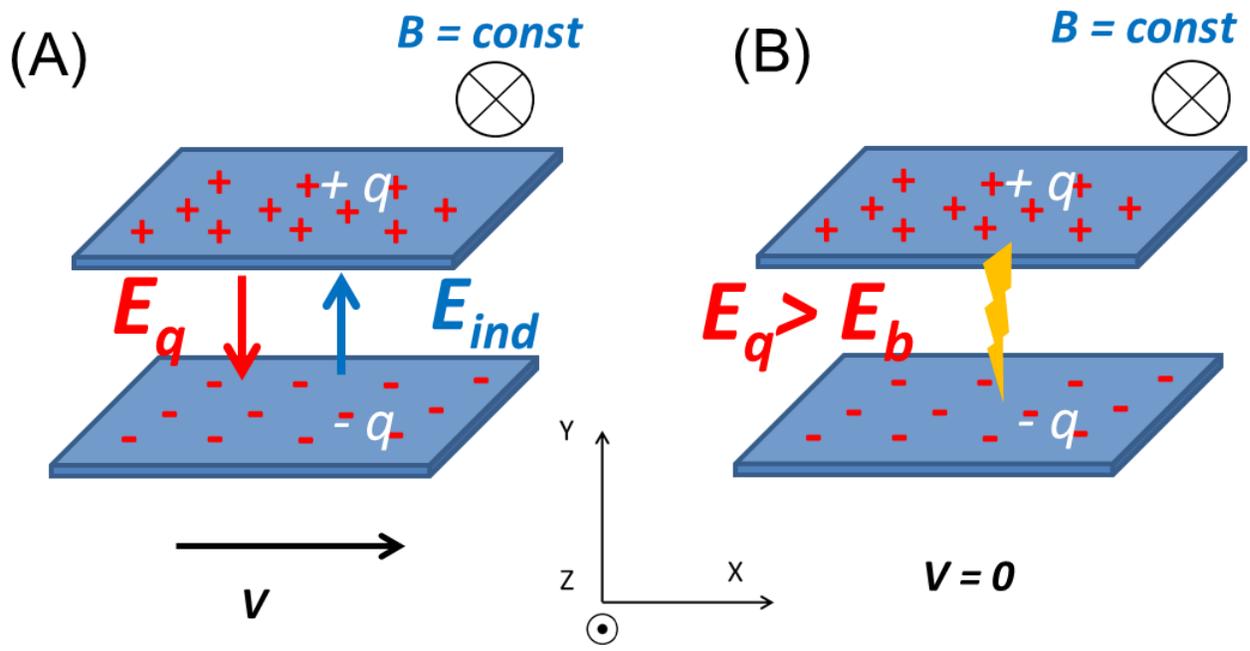

**Figure 2**

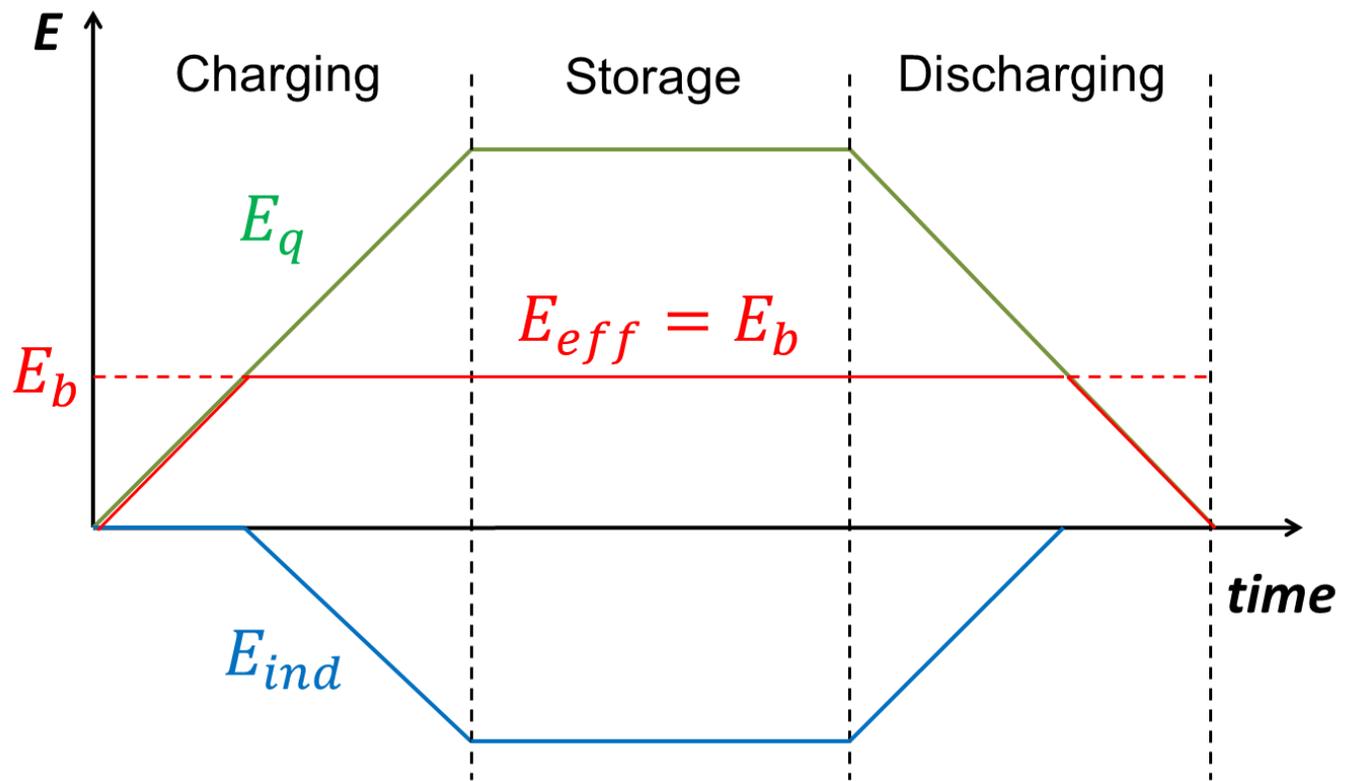

**Figure 3**

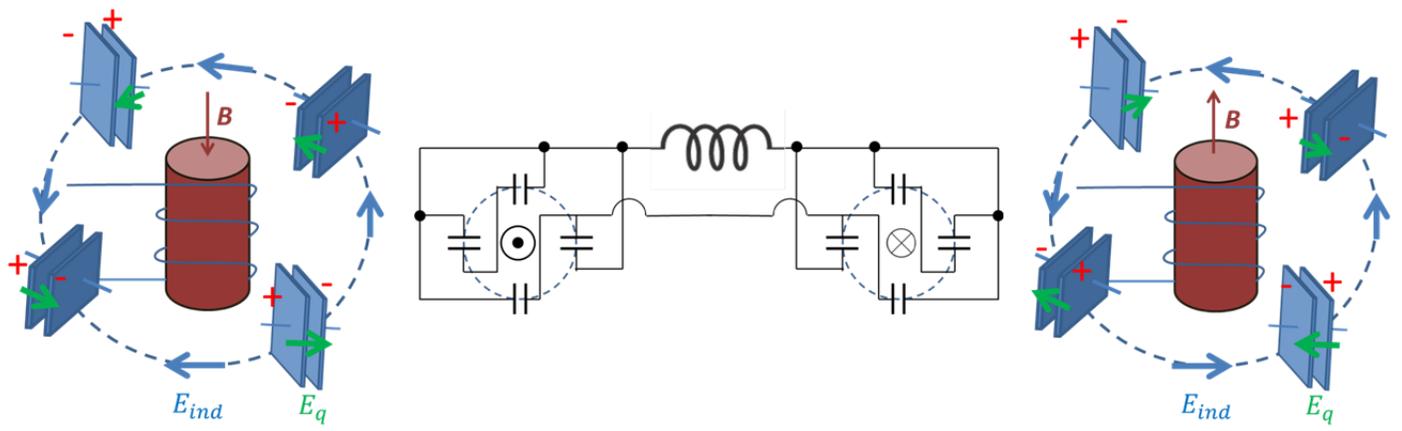

**Figure 4**

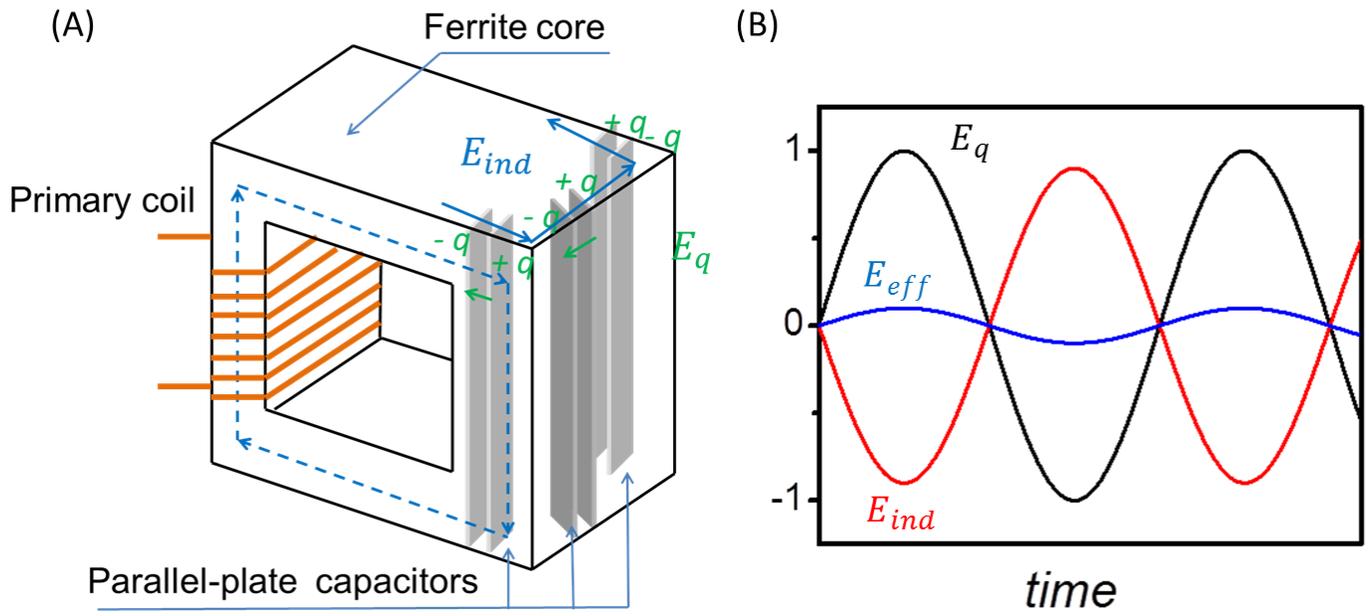

**Figure 5**

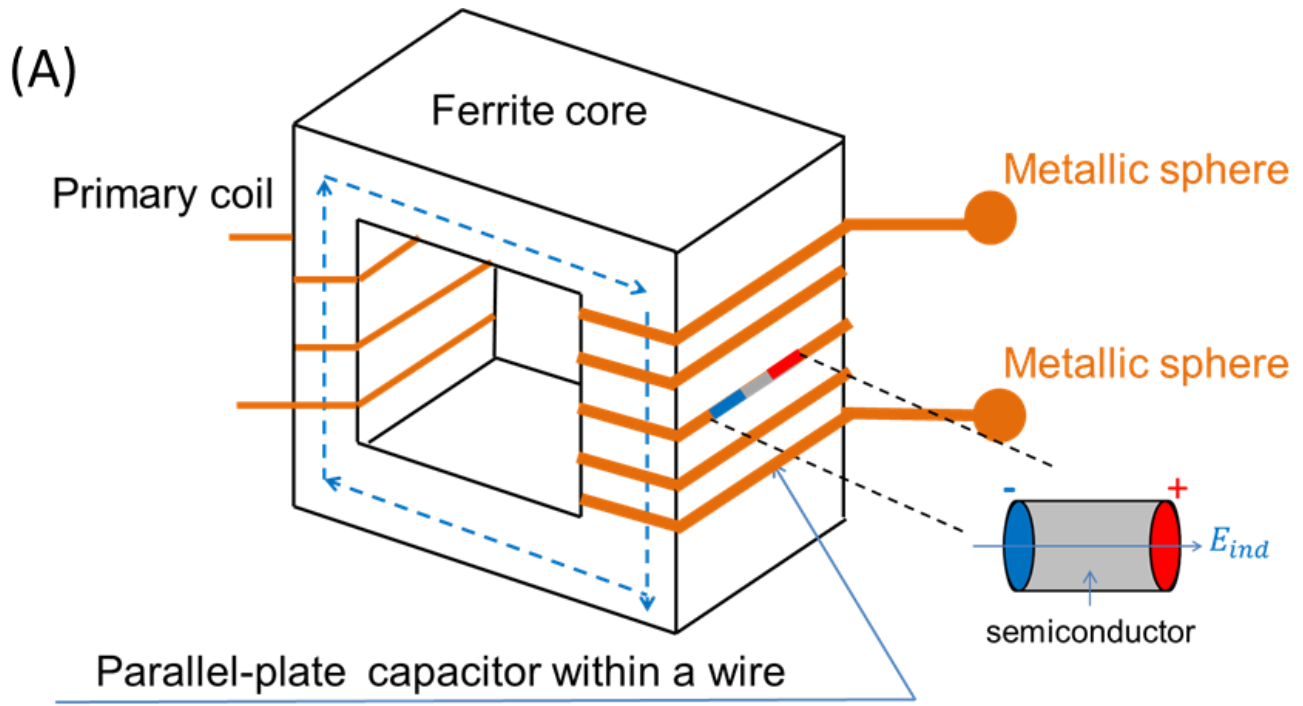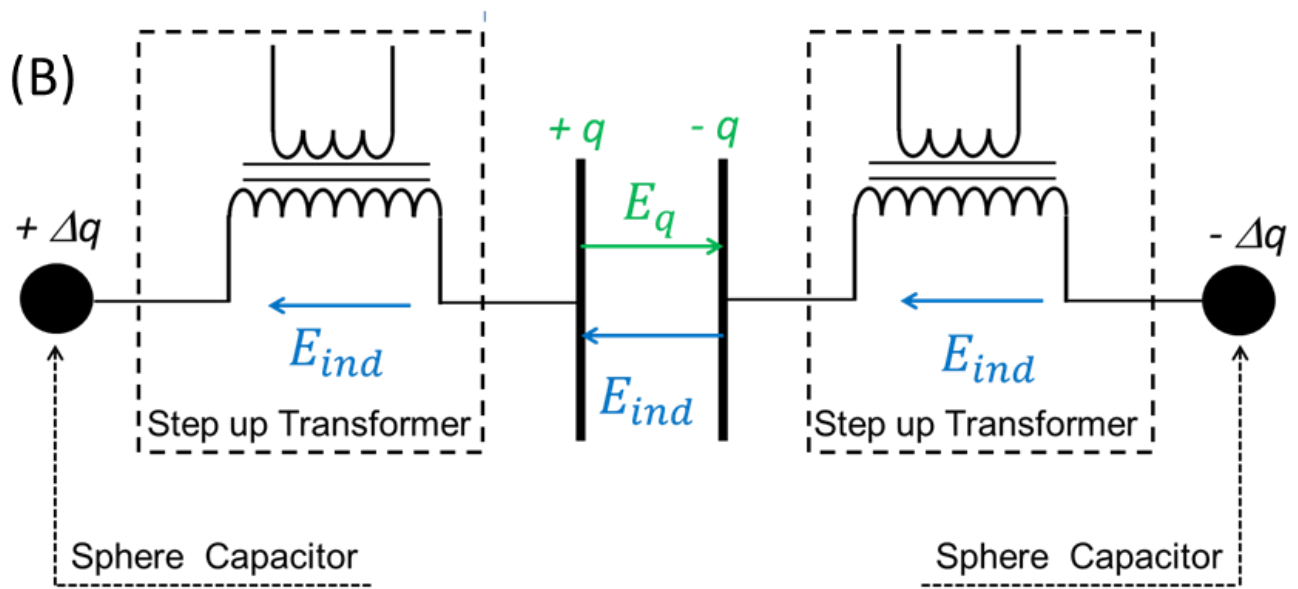

**Figure 6**